# Split ring resonator-coupled enhanced transmission through a single subwavelength aperture

Koray Aydin,[1] A. Ozgur Cakmak,[1] Levent Sahin,[1] Z. Li,[1] Filiberto Bilotti,[2] Lucio Vegni,[2] and Ekmel Ozbay[1]

[1]*Department of Physics, Nanotechnology Research Center and Department of Electrical and Electronics Engineering, Bilkent University, 06800, Ankara, TURKEY*

[2]*Department of Applied Electronics, University of Roma Tre, Rome 00146, Italy*

**Abstract**

We report the enhanced transmission of electromagnetic waves through a single subwavelength aperture by making use of the resonance behavior of a split ring resonator (SRR) at microwave frequencies. By placing a single SRR at the near-field of the aperture, strongly localized electromagnetic fields are effectively coupled to the aperture with a radius that is twenty times smaller than the resonance wavelength ($r/\lambda = 0.05$). We obtained 740-fold transmission enhancement by exciting the electric resonance of SRR. A different coupling mechanism, through the magnetic resonance of SRR, is also verified to yield enhanced transmission. Good agreement is obtained between the microwave measurements and numerical simulations.

Enhanced transmission of light through apertures that are much smaller than the wavelength has received a burgeoning amount of interest [1,2], after the seminal work by Ebbesen *et al.,* realizing the extraordinary transmission from subwavelength hole arrays that were milled in optically thin metallic films [3]. Extraordinary light transmission has been extensively studied by using subwavelength periodic hole arrays [3-7] or metallic structures with a single aperture [7-12]. Based on Bethe's theoretical description, transmission through a single subwavelength aperture of a radius $r << \lambda$ scales with $(r/\lambda)^4$ [13]. However, one can increase the amount of light passing through a single hole by corrugating the metal surface with periodic grooves [7], by filling the hole with a material of high dielectric permittivity [8,9], by using alternative aperture geometries [10,11] or by placing artificially designed metamaterial covers in front of the aperture [12]. In all of the approaches, transmission is enhanced by a resonant process that leads to the effective coupling of light to a small aperture. Although much of the work on enhanced transmission has been carried out at optical frequencies, similar results are also obtained at microwave [14,15] and THz [16] frequency regimes.

In this Letter, we propose and demonstrate an alternative approach that utilizes the resonance of a split-ring resonator (SRR) [17] in order to enhance the transmission through a single subwavelength aperture at microwave frequencies. We successfully demonstrated the extraordinary transmission of microwave radiation through an aperture of radius $r$, which is twenty times smaller than the incident wavelength ($r/\lambda = 0.05$). To our knowledge, this is the smallest aperture size utilized for achieving enhanced transmission. We measured a 740-fold enhancement by using electrically coupled SRR in the proximity of a single aperture. However, it is also possible to enhance the transmission by exciting the magnetic resonance of SRRs. Enhanced transmission is attributed to the highly localized electric fields at the split region of SRRs that couple incident EM radiation to the aperture around SRR's resonance frequency.

Figure 1(a) shows the schematics of the structure under investigation. A commercial 1.6 mm thick FR4 printed circuit board deposited with a thin (30 μm) copper plate of the size $L \times L$ ($L = 200$ mm) was used in the experiments. A circular aperture with a radius of $r = 4$ mm at the center of a metallic plate was created by mechanical etching. In the experiments, a split-ring resonator was used to couple electromagnetic (EM) waves to the aperture at a resonance frequency of SRR. The split ring resonator was comprised of two concentric rings with the parameters as provided in [18]. The radius of the SRR is 3.6 mm, which is comparable with the aperture size. Figure 1(b) shows an alternative configuration of placing SRR in front of the aperture, wherein a different resonance mechanism (magnetic resonance) plays a role in the enhanced transmission.

The resonance of a single SRR was verified simply by measuring the transmission spectrum. Transmission measurements were performed by using two waveguides as a transmitter and a receiver, which were connected to the Agilent N5230A portable network analyzer. Four different orientations of SRR with respect to the incident EM wave were investigated, in which the results are shown in Fig. 1(c). For the configurations SRR (A) [$E_{y'}$, $H_{x'}$] and SRR (B) [$E_{x'}$, $H_{y'}$], the wave propagation was perpendicular to the plane of SRRs [$k_{z'}$] (see inset for the directions). Although magnetic resonance could not be excited, it is possible to excite an electrical resonance for SRR (A), due to the asymmetry of the SRR structure with respect to the electric field [16]. The resonance frequency for SRR (A) is measured at 3.55 GHz. For the SRR (B) case, the transmission resonance was not observed. The magnetic resonance in SRRs could only be excited when $H$ is perpendicular to the plane of SRR as in the cases of SRR (C) [$k_{x'}$, $E_{y'}$] and SRR (D) [$k_{y'}$, $E_{x'}$]. Both electric and magnetic fields contribute to the resonance mechanism of SRRs for configuration (C) [19]. The resonance of SRR (D) is purely of magnetic origin since $E$-field is symmetric with the orientation of the splits. The resonance frequencies were 3.82 GHz for SRR (C) and 3.85 GHz for SRR (D).

In the measurements, we employed two waveguide antennae to transmit and receive electromagnetic waves. A transmitter antenna was placed 0.2 mm away from the metallic plate with the aperture and the transmitted power was collected 5 cm away from the structure. Wave propagation was along the $z$ axis, where $E // y$ and $H // x$ (Fig. 1(a)). We first measured the transmission through the reference sample, which was the metallic plate with a single subwavelength aperture (red dashed line in Fig. 2(a)). As expected from the diffraction theory, the transmission of electromagnetic waves through a small aperture was very weak [13]. We then placed a single SRR, 0.1 mm away from the aperture with the outer split region facing the center of the aperture as shown in Fig. 1(a). The blue line in Fig. 2(a) plots the measured intensity of the EM wave, which propagated through the SRR (A) structure and aperture. The transmission is significantly increased when a single SRR was placed at the near-field of the aperture. We observed a 740-fold enhancement in the transmission at 3.55 GHz as shown in the inset. Here, we define the enhancement as the ratio of the field intensity of a transmitted EM wave through an SRR and aperture to that through the aperture only. It is noteworthy that the maximum enhancement was obtained at the resonance frequency of SRR (A). The electrically excited resonance of SRR causes the strong localization of the electric field at the splits and gaps of an SRR structure.

The resonance of SRR is responsible for the enhancement in the transmission. In order to verify this evidence, we performed additional measurements. We placed a closed ring resonator (CRR), in which the structure comprised of two concentric rings without splits [18]. Evidently, the resonance behavior was no longer present for CRR and the transmission was comparable with that of the aperture-only structure. We also checked the SRR (B) configuration together with the aperture, for which *E // x* and *H // y*. Since the resonant coupling of EM waves to the aperture is not present for this configuration, transmission is not enhanced through the aperture. In our approach, where we utilized a bianisotropic SRR, the enhancement of microwave radiation depends on the polarization of the incident EM wave. It has been previously shown that apertures with rectangular [10,11] and elliptical [6] shapes yield enhanced transmission for a specific polarization of an incident EM wave. Here, we achieved strong polarization dependence of enhanced transmission through a circular aperture by using a bianisotropic SRR structure.

Numerical simulations were performed by using the commercially available software, CST Microwave Studio. The dielectric constant and tangent loss of the FR4 dielectric substrate were taken as $\varepsilon = 3.6$ and $\delta = 0.01$, respectively. We modeled the waveguide antennae that were used in the experiments to transmit and detect the electromagnetic waves. Open boundary conditions were applied along all the directions. Figure 2(b) shows the numerical simulation results for only aperture, CRR, SRR (A) and SRR (B) configurations. There was a good agreement between the measured and simulated results. In the simulations, the maximum transmission through SRR (A) and aperture is observed at 3.70 GHz and the enhancement factor is calculated as 820. The differences between the measured and simulated results were attributed to the deviation from the ideal material parameters and the sensitivity of the distance between the SRR and aperture on the transmission.

One can also achieve enhanced transmission through SRR and aperture by exciting the magnetic resonance of SRR. A single SRR is placed in front of the aperture perpendicularly, as is schematically drawn in Fig. 1(b). SRR (D) is symmetric with respect to the $E$-field, whereas SRR (C) is antisymmetric. The corresponding results are plotted in Fig. 3(a) and Fig. 3(b). The orientations of SRR are shown in the relevant figure insets. Measurements (solid blue line) and simulations (dashed dotted line) revealed that the transmission through the subwavelength aperture was increased for both SRR configurations. Configuration (D) yielded a higher transmission compared to that of configuration (C). The magnetic field excited the resonance in both orientations. However, due to the asymmetry of the SRR (C) structure with respect to the incident $E$-field, there occurred an electric coupling to the magnetic resonance, which was not present in SRR (D). Obviously, additional electric resonance hinders the performance of enhanced transmission through SRR and aperture. Enhancement was measured to be 38 for configuration (D) at 3.84 GHz and 7.5 for configuration (C) at 3.76 GHz as plotted in Fig. 3(c). The simulation results were consistent with the experiments where we calculated the enhancement factors of 34 and 7.8 for (D) and (C) configurations.

In order to gain insight to the physical mechanisms behind the transmission enhancement processes, we performed additional numerical simulations to evaluate the amplitude of the electric field at frequencies where our simulations predicted the highest transmission. Figure 4(a) plots the $E_y$ amplitude evaluated at $y=0$ plane, corresponding to the center of the structure. The amplitude of the electric field of the incident plane wave was unity. The wave was incident from –z and propagated along the +z direction. The aperture was located at z=0, between x/r = -1 and x/r = 1. In the figures, the x and z values were given in terms of the radius of the aperture r. The solid rectangles show the positions of the dielectric FR4 substrates on which the SRR and metallic plate were deposited. The fields are localized strongly at the split regions of the SRR. The coupling of fields to the aperture via SRR is clearly seen. Figure 4(b) shows the $E_y$ amplitude for the SRR (D) configuration that is evaluated at y=3 plane, where the outer split of the SRR is located. In this case, the incident EM wave was coupled through the aperture via the magnetic resonance of SRR. The amplitude of $E_y$ for the SRR (C) and aperture at y=0 plane is shown in Fig. 4(c). Although the field enhancement is higher for the SRR (C) structure compared to that of SRR (D), the transmitted electromagnetic wave is apparently smaller.

In all of the experiments and simulations, the aperture radius was 20 times smaller than the wavelength ($r/\lambda = 0.05$), whereas the typical aperture sizes utilized to achieve enhanced transmission was $r/\lambda = 0.20$ [3]. We have been able to enhance the transmission of EM waves through a single aperture that was four times smaller than the size of conventional subwavelength apertures. The reduction of aperture size is achieved by employing a single resonant element that resonates at much longer wavelengths compared to its size (~$\lambda/10$). Since the frequency of an enhanced transmission was determined by the resonance frequency of SRR, one can employ an active SRR medium [20] to tune the wavelength of maximum enhancement. Utilizing SRR also provides design flexibility, in which one can easily fabricate SRRs to resonate at a desired frequency.

Our experiments were carried out at microwave frequencies, but similar results could also be obtained from THz [21] to near IR [22] frequencies, since SRRs have already been realized at these frequency regimes. To achieve an enhanced transmission at optical frequencies, one can fabricate an optical antenna close to a subwavelength aperture. The fields were strongly localized in the antenna feed [23], similar to the localization of fields at the SRRs split region. It is noteworthy that the metals are perfect conductors at microwave frequencies and, therefore, surface plasmons do not contribute to the enhancement process. Furthermore, surface waves were not present in our approach, since we did not use a grating structure to enhance the radiation. The strongly localized fields around the SRR couple the EM waves through a resonant process to the subwavelength aperture.

To conclude, we successfully demonstrated the enhanced transmission of microwave radiation through a single subwavelength aperture by the placement of a single resonant element in the proximity of the aperture. The enhanced transmission was achieved by exciting the electric and/or magnetic resonance of SRR. A 740-fold enhancement was obtained through an aperture that was twenty times smaller than the wavelength. We also found that there was strong polarization dependence for enhancing the transmission, due to the bianisotropic nature of a split-ring resonator.

This work is supported by the European Union under the projects EU-METAMORPHOSE, EU-PHOREMOST, EU-PHOME, and EU-ECONAM, as well as TUBITAK under the Project Numbers 105E066, 105A005, 106E198, and 106A017. One of the authors (E.O.) also acknowledges partial support from the Turkish Academy of Sciences.

**REFERENCES**

1. C. Genet, and T. W. Ebbesen, Nature **445**, 39 (2007).
2. F. J. Garcia de Abajo, Rev. Mod. Phys. **79**, 1267 (2007).
3. T. W. Ebbesen *et al.*, Nature **391**, 667 (1998).
4. L. Martin-Moreno *et al.*, Phys. Rev. Lett. **86**, 1114 (2001).
5. W. L. Barnes *et al.*, Phys. Rev. Lett. **92**, 107401 (2004).
6. R. Gordon *et al.*, Phys. Rev. Lett. **92**, 037401 (2004).
7. H. J. Lezec *et al.*, Science **297**, 820 (2002).
8. F. J. Garcia de Abajo, Opt. Express **10**, 1475 (2002).
9. K. J. Webb, and J. Li, Phys. Rev. B **73**, 033401 (2006).
10. A. Degiron *et al.*, Opt. Comm. **239**, 61 (2004).
11. F. J. Garcia-Vidal *et al.*, Phys. Rev. Lett. **95**, 103901 (2005).
12. A. Alu *et al.*, IEEE Trans. Antennas Propag. **54**, 1632 (2006).
13. H. A. Bethe, Phys. Rev. **66**, 163 (1944).
14. S. S. Akarca-Biyikli, I. Bulu and E. Ozbay, Appl. Phys. Lett. **85**, 1098 (2004).
15. H. Caglayan, I. Bulu, and E. Ozbay, Optics Express **13**, 1666 (2005).
16. Gomez-Rivas *et al.*, Phys. Rev. B 68, 201306 (2003).
17. J. B. Pendry *et al.*, IEEE Trans. Microwave Theory Tech. **47**, 2075 (1999).
18. K. Aydin *et al.*, Opt. Lett. **29**, 2623 (2004).
19. N. Katsarakis *et al.*, Appl. Phys. Lett. **84**, 2943 (2004).
20. H.-T. Chen *et al.*, Nature **444**, 597 (2006).
21. T. J. Yen *et al.*, Science **303**, 1494 (2004).
22. C. Enkrich et al., Phys. Rev. Lett. **95**, 203901 (2005).
23. P. Muhlschlegel et al., Science **308**, 1607 (2005).

**FIGURE CAPTIONS**

FIG. 1. (color online) Schematic drawings of a subwavelength aperture in a metallic plate and SRR plane (a) parallel, and (b) perpendicular to the aperture plane. (c) Measured transmission spectra from a single SRR with four different orientations.

FIG. 2. (color online) (a) Measured and (b) simulated intensity of transmitted EM wave from only aperture, and aperture covered with CRR, SRR (A) and SRR (B). The insets show the enhancement factor obtained from SRR (A) and aperture configuration.

FIG. 3. (color online) Measured and simulated intensity of transmitted EM waves from a single aperture covered with (a) SRR (D) and (b) SRR (C). (c) Enhancement factors obtained from the measurements and simulations for configurations (C) and (D).

FIG. 4. (color online) Simulated $E_y$ magnitude for an aperture and (a) SRR (A), (b) SRR (D) and (c) SRR (C) structure.

**LIST OF FIGURES**

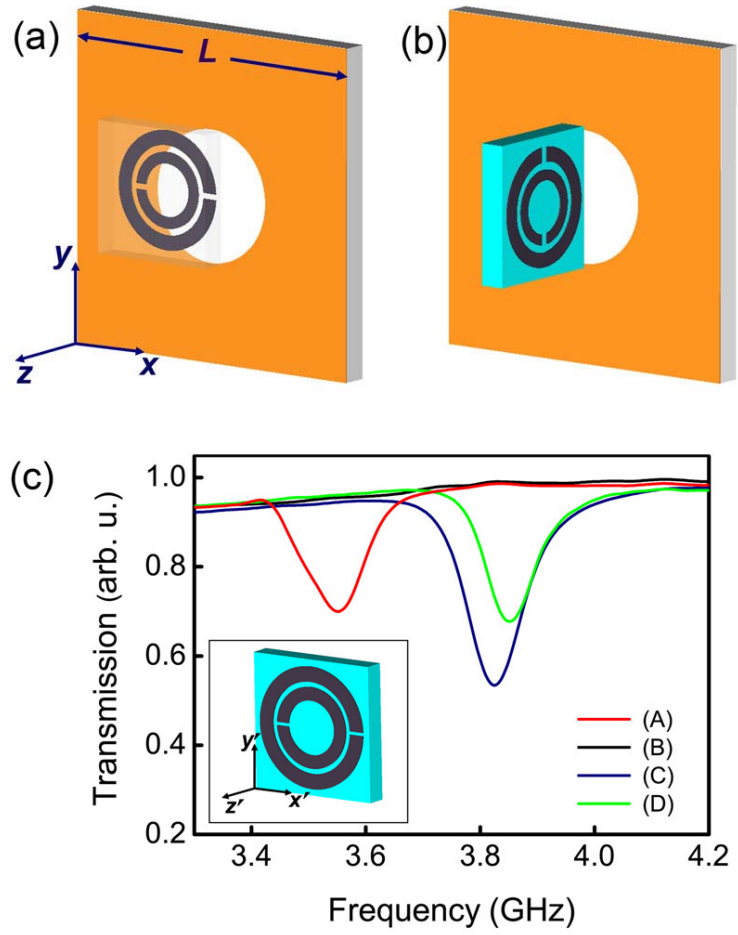

**FIG. 1**

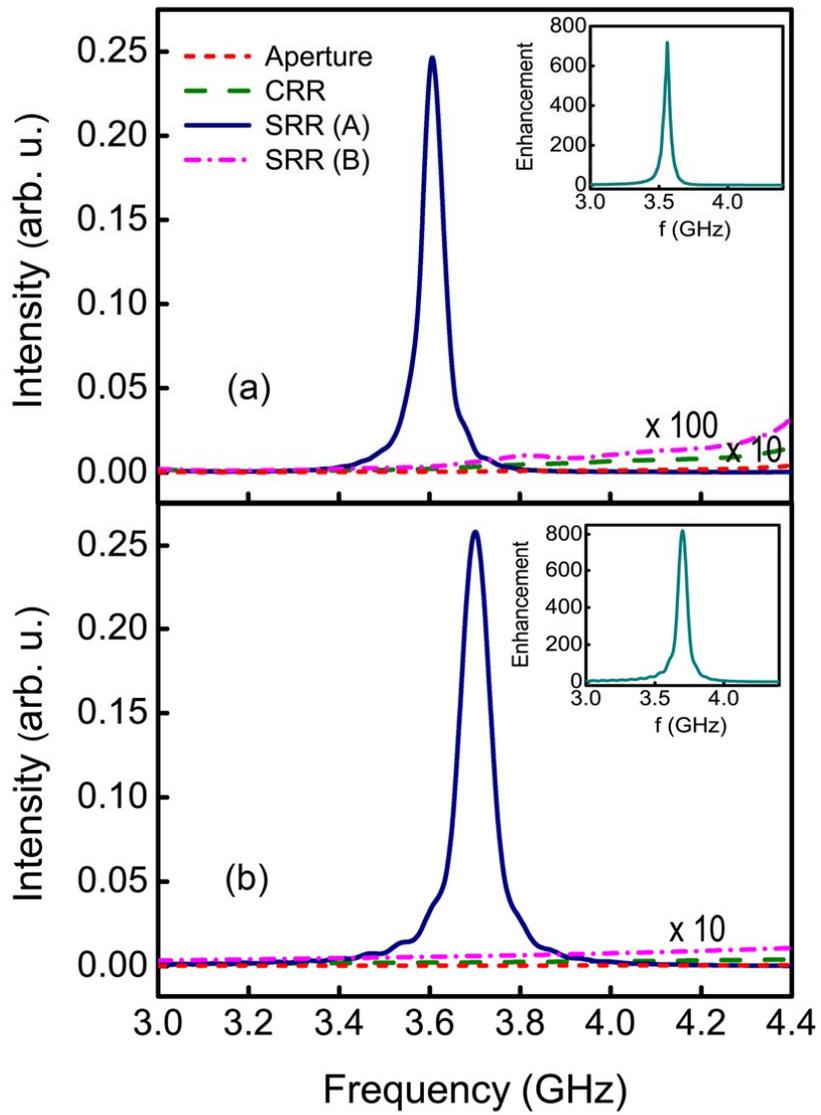

**FIG. 2**

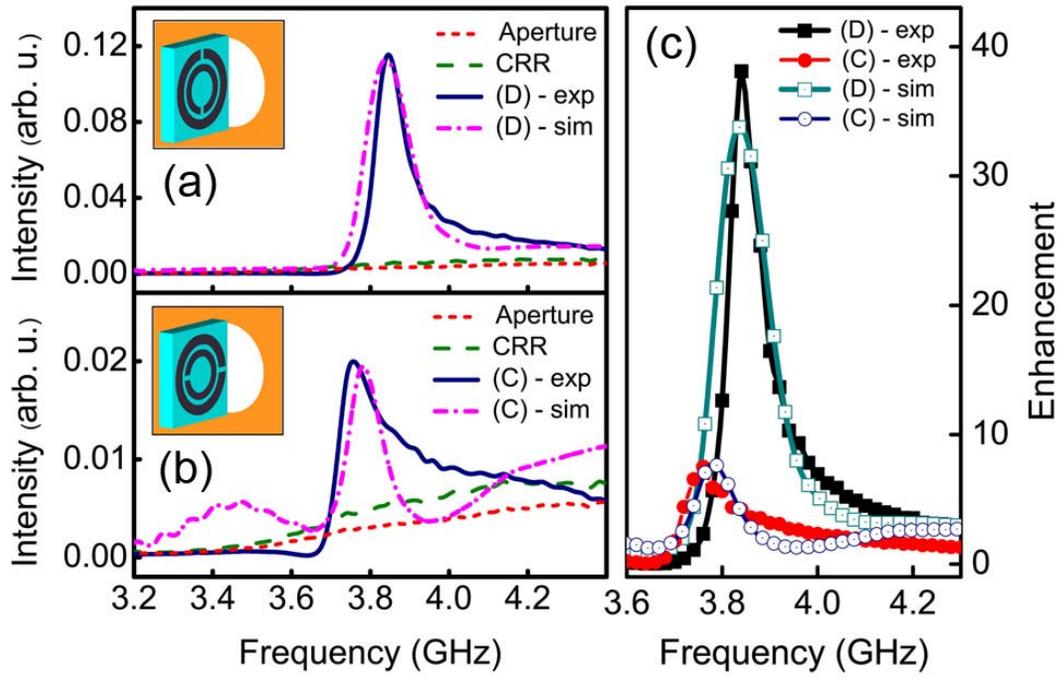

**FIG. 3**

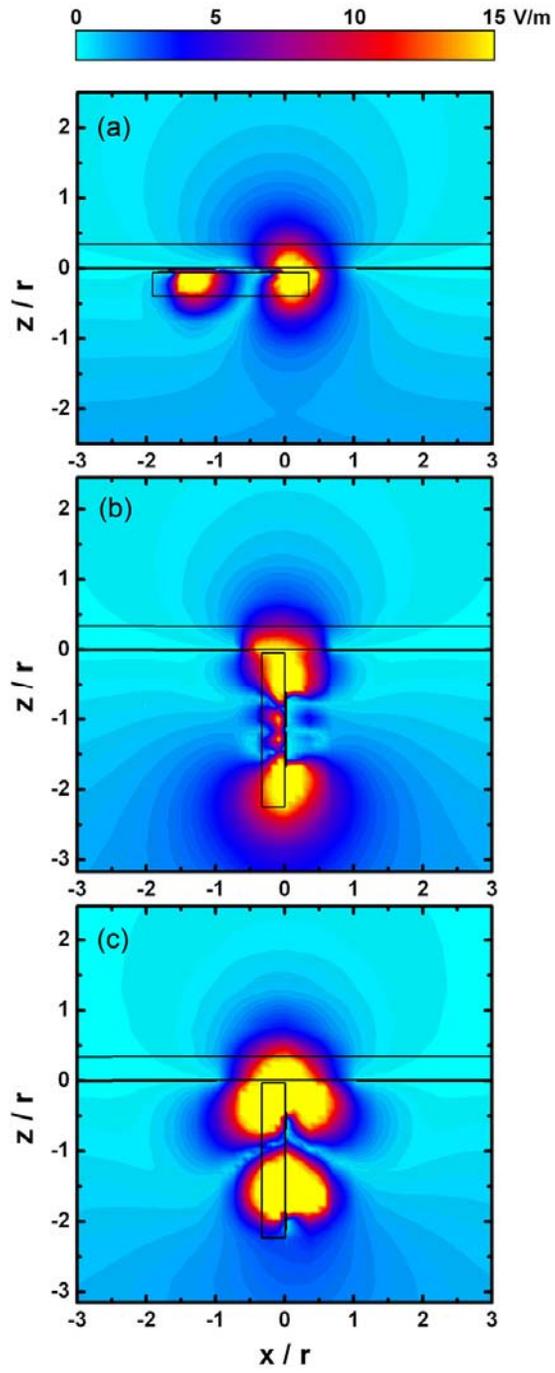

**FIG. 4**